\documentclass[12pt]{iopart}
\usepackage{iopams,bbm,cite}

\newcommand{\average}[1]{\langle#1\rangle}
\newcommand{\bigaverage}[1]{\Big\langle#1\Big\rangle}
\newcommand{\dd}[2]{\frac{\partial #1}{\partial #2}}
\newcommand{\ddc}[3]{\left(\dd{#1}{#2}\right)_{\!#3}}
\providecommand{\eqref}[1]{(\ref{#1})}

\begin{document}

\title{Green's functions and hydrodynamics for isotopic binary diffusion}

\author{R. van Zon and E. G. D. Cohen}

\address{The Rockefeller University, 1230 York Avenue, New York, NY 10021-6399, USA}

\begin{abstract}
   We study classical binary fluid mixtures in which densities vary on
   very short time (ps) and length (nm) scales, such that
   hydrodynamics does not apply. In a pure fluid with a localized heat
   pulse the breakdown of hydrodynamics was overcome using Green's
   functions which connect the initial densities to those at later
   times.  Numerically it appeared that for long times the results
   from the Green's functions would approach hydrodynamics.  In this
   paper we extend the Green's functions theory to binary mixtures.
   For the case of isothermal isobaric mutual diffusion in isotopic
   binary mixtures and ideal binary mixtures, which is easier to
   handle than heat conduction yet still non-trivial, we show
   analytically that in the Green's function approach one recovers
   hydrodynamic behaviour at long time scales provided the system
   reaches local equilibrium at long times. This is a first step
   toward giving the Green's function theory a firmer basis because it
   can for this case be considered as an extension of hydrodynamics.
\end{abstract}

{ 
  \footnotesize
  \noindent
  \hspace{6pc}\textit{Dated 10 August 2005}

  \pacs{47.10.+g, 05.20.Jj, 66.10.Cb, 05.60.Cd}

  \begin{tabbing}
  \hspace{6pc}KEYWORDS:\hspace{2.7mm}
  \= Transport properties~(theory), dynamical processes~(theory), \\
  \> dynamical heterogeneities~(theory), binary mixtures, \\
  \> isotopes, Green's functions, hydrodynamics
  \end{tabbing}

}

\maketitle

\tableofcontents\pagestyle{plain}

\section{Introduction}
\label{introduction}

With the increase in miniaturization nanoscale technology is
developing rapidly. However, at such small scales the physics of
thermodynamic properties and transport processes will deviate from
that at larger scales. Certainly one sees finite size effects and
appreciable fluctuations on measurable quantities. It is likely that
new physics exists at the nanoscale.
                              
Of course, on large time and length scales transport processes in
fluids are well described by hydrodynamics which contain the transport
coefficients such as the diffusion constants, viscosity and thermal
conductivity \cite{DeGrootMazur}.  Kinetic theory can provide low
density approximations to these coefficients \cite{ChapmanCowling}
while linear response theory yields the Green-Kubo relations which
express the transport coefficients in terms of integrals of
time-correlation functions of the appropriate current fluctuations in
equilibrium \cite{Wood89,Steele}. Thus once one has determined the
transport coefficients by studying the equilibrium time-correlation
functions, one can attack a wide range of non-equilibrium transport
problems described by hydrodynamics not too far from equilibrium. In
contrast, the description of the non-equilibrium behaviour at shorter
times and smaller length scales remains mostly unsolved despite its
importance, in principle, to the understanding of nanostructures. Even
low density kinetic theories cannot describe time scales below the
average duration of a collision.  Attempts to generalize hydrodynamics
stranded at first because a straightforward further expansion in the
gradients, i.e.\ beyond the linear laws, leads to higher order Burnett
coefficients which are infinite \cite{DorfmanCohen67}. Recently a
theory was proposed by Prof.~J.M.~Kincaid using Green's functions to
describe processes at small length and time scales which avoids these
divergences
\cite{Kincaid95,KincaidCohen02a,KincaidCohen02b,CohenKincaid02}.  Like
hydrodynamics the Green's function approach can also treat the
hydrodynamic fields (density, momentum and energy) but unlike in
hydrodynamics currents need not be linear in the gradients.  Another
contrast with hydrodynamics is that the Green's function approach is
not restricted to long times.  Thus the Green's function approach
should be able to attack a wider range of problems than standard
hydrodynamics.  Indeed the Green's function approach has been
numerically applied with success to self-diffusion \cite{Kincaid95}
and heat transport
\cite{KincaidCohen02a,KincaidCohen02b,CohenKincaid02} on nanometre\
and picosecond scales.  In these applications, the Green's functions
are resummed into an infinite series that needs to be truncated.  It
was shown using computer simulations that if just a few terms from the
infinite sum are taken into account, accurate results were obtained
for all time scales for those systems
\cite{Kincaid95,KincaidCohen02a,KincaidCohen02b,CohenKincaid02}.
Also, for long times the numerical results were seen to approach the
numerical hydrodynamic results \cite{CohenKincaid02}.

In this paper we formulate the Green's function approach for mixtures,
focusing on \emph{mutual diffusion} in isotopic and ideal binary
mixtures. Apart from adapting the Green's function approach for mass
transport in mixtures, this paper is mainly dedicated to the question
whether the truncated expressions for the Green's functions describe
the right very short time behaviour and whether they do indeed reduce
to hydrodynamics at long time scales and for small gradients; if the
latter is true then the Green's function approach can rightfully be
called an extension of hydrodynamics.  For self-diffusion in a single
component fluid the truncation was already shown analytically to give
the right hydrodynamic-like behaviour for long times
\cite{Kincaid95}. Mutual diffusion in an isotopic or ideal mixture is
the simplest case to consider next (simplest both in thermodynamic and
in hydrodynamic aspects) Ideally one would like to show analytically
that the Green's functions reproduce not just mutual diffusion but the
full hydrodynamic equations in all cases where gradients are small and
times are long, but this would be an enormous undertaking. Therefore,
as a first step, we restrict ourselves in this paper to the already
nontrivial case of diffusion in isotopic and ideal mixtures.

\section{Classical isotopic and ideal binary mixtures}
\label{section2}

\subsection{Definitions}
\label{class-binary-mixt}

We consider a classical mixture of two components where the particles
interact through short-range pair potentials. The total system
consists of $N$ particles of which $N_1$ particles are of component
$1$ and $N_2=N-N_1$ particles are of component $2$. The system is
enclosed in a volume $V$. We denote the (three-dimensional) position
of particle $ i$ of component $\lambda$ as $\bi r_{\lambda i}$ and its
velocity as $\bi v_{\lambda i}$, where the index $\lambda$ equals 1 or
2, and $ i$ is an index which runs from 1 to $N_\lambda$.  The mass of
particles of component $\lambda$ will be denoted as $m_\lambda$. We
restrict ourselves to two related classes of mixtures: isotopic
mixtures, for which the pair potentials are the same independent of
the components to which the interacting particles belong, and ideal
mixtures, whose pair potentials can be neglected as far as
thermodynamic properties are concerned.

The system is prepared in a non-equilibrium state described by some
given, non-equilibrium initial ensemble in phase space $\mathcal
P_{\mathrm{ne}}(\Gamma)$, where $\Gamma$ $=$ $\{\bi r_{\lambda i}, \bi
v_{\lambda i}\}$.  In the initial non-equilibrium ensemble --- and
consequently also at later times --- the particles of each component
are indistinguishable.  After the initial preparation in the
non-equilibrium state, the system stays isolated, i.e.\ the system
does not interact with the outside.

The macroscopic (i.e.\ averaged) local density of component $\lambda$
is defined as \cite{IrvingKirkwood50}
\begin{eqnarray}
  n_\lambda(\bi r, t)
&\equiv&
  \int\! d\Gamma\, \mathcal P_{\mathrm{ne}}(\Gamma) \sum_{ i=1}^{N_\lambda} 
    \delta\mathbf(\bi r-\bi r_{\lambda  i}(t)\mathbf)
=
N_\lambda  
\average{\delta\mathbf(\bi r-\bi r_{\lambda1}(t)\mathbf)}_{\mathrm{ne}}
\label{1.1}
\end{eqnarray}
where the subscript ``$\mathrm{ne}$'' on the average indicates its
non-equilibrium nature.  In this equation we used the
indistinguishability of the particles of the same component to express
the densities $n_\lambda(\bi r,t)$ in terms of the properties of a
representative particle of each component i.e.\ in terms of $\bi
r_{\lambda1}(t)$.

Due to the non-equilibrium nature of $\mathcal P_{\mathrm{ne}}$, the
local densities $n_\lambda(\bi r,t)$ will in general not be uniform at
$t=0$.  In this paper we are mainly interested in the time evolution
of these densities.  This is in principle determined by the
microscopic time evolution governed by the Hamiltonian
\begin{equation}
  H(\Gamma) = \sum_{\lambda=1}^2
     \sum_{ i=1}^{N_\lambda} \case12m_\lambda|\bi
  v_{\lambda  i}|^2 + 
   U(\Gamma)
\label{Hamiltonian}
\end{equation}
where
\begin{equation}
U(\Gamma)=
  {\sum_{\lambda,\nu, i, j}}{\Big.}'\case12 
\varphi_{\lambda\nu}(|\bi r_{\lambda  i}-\bi r_{\nu j}|).
\label{Udef}
\end{equation}
Here $\varphi_{\lambda\nu}$ are the pair potentials and the summation
$\sum{\Big.}'$ is restricted to exclude self-interaction of the
particles, i..e.\ to exclude the cases where both $\lambda=\nu$ and
$i=j$.  For isotopic mixtures,
$\varphi_{12}=\varphi_{11}=\varphi_{22}$. The equations of motion are
thus given by
\begin{eqnarray}
  \dot{\bi r}_{\lambda i} &=& \bi v_{\lambda i}
\label{rdot}
 \\ 
  \dot{\bi v}_{\lambda i} &=& 
- \frac{1}{m_\lambda}\dd{U}{\bi r_{\lambda i}}.
\label{vdot}
\end{eqnarray}

Before proceeding we introduce a few more quantities for later
convenience.  The local mass densities are given by $\rho_\lambda =
m_\lambda n_\lambda$.  Here and below when the spatial and temporal
arguments of a local quantity are not denoted, it is implied that they
are taken at position $\bi r$ and time $t$.  The local total number
and mass densities are, respectively, $n = n_1 + n_2$ and $\rho =
\rho_1 + \rho_2$.  We furthermore define the local mole fractions as
$x_\lambda = n_\lambda /n$ and the mass fraction of component $1$ as
$c = \rho_1 /\rho$.  The mass fraction of component $2$ is simply
$1-c$.  The local mass currents of the two components in the
laboratory frame are given by ($\lambda=1,$ $2$)
\begin{eqnarray}
  \bi j^*_\lambda  &=& N_\lambda m_\lambda \average{\bi v_{\lambda1}(t) 
    \delta\mathbf(\bi r-\bi r_{\lambda1}(t)  \mathbf)}_{\mathrm{ne}}
\label{1.8}
\end{eqnarray}
Using these definitions the local mass velocity of the fluid can be
written as
\begin{equation}
 \bi u = (\bi j_1^* +\bi j_2^* )/\rho.
\label{udef}
\end{equation}
Finally, we define the diffusive currents of the two components in the
local frame (i.e.\ co-moving with the local velocity $\bi u$) as
\begin{eqnarray}
\label{1.11}
  \bi j_\lambda
&=&
 \bi j_\lambda^* - \rho_\lambda\bi u
=
 N_\lambda m_\lambda\average{[\bi v_{\lambda1}(t)-\bi u]
\,\delta\mathbf(
\bi r-\bi r_{\lambda1}(t)  \mathbf)}_{\mathrm{ne}}
\end{eqnarray}
so that by definition we have
\begin{equation}
\bi j_1+\bi j_2 = 0.
\label{1.12}
\end{equation}

\subsection{Thermodynamics, hydrodynamics and Green-Kubo formula}
\label{hydr-green-kubo}

For classical isotopic mixtures and ideal mixtures, many thermodynamic
equilibrium properties take on a particularly simple form --- which is
in fact the main reason we consider these systems here.  In
particular, for the current case of mass transport investigated in
this paper, we will need $(\partial\mu/\partial c)_{Tp}$ and
$(\partial\rho/\partial c)_{Tp}$ where $T$ is the temperature, $p$ is
the pressure and
\begin{equation}
\mu=\frac{\mu_1}{m_1}-\frac{\mu_2}{m_2}
\label{mudef}
\end{equation}
and $\mu_\lambda$ is the chemical potential of component $\lambda$.
These thermodynamic properties can be found from the form of chemical
potential for isotopic systems \cite[\S 93]{LandauLifshitzSP}:
\begin{eqnarray}
\mu_\lambda = k_BT\log x_\lambda + \mu_{0\lambda};\quad
\label{muiiso}
\mu_{01}  = \mu_{02}  -\case32 k_BT \log\frac{m_1}{m_2},
\end{eqnarray}
where $\mu_{0\lambda}$ is the chemical potential of the pure isotope
$\lambda$ at the same temperature $T$ and pressure $p$ as those for
which $\mu_\lambda$ is evaluated. Equation \eqref{muiiso} implies that
\begin{eqnarray}
  \ddc{\mu_\lambda}{x_\lambda}{Tp} = \frac{k_BT}{x_\lambda}
\label{firstrelation}
;\quad
  \ddc{\mu_1}{p}{Tx} =   \ddc{\mu_2}{p}{Tx}.
\end{eqnarray}
We can use the Gibbs-Duhem relation\cite[\S 8.7]{Reif}
\begin{equation}
 N_1 d\mu_1 + N_2 d\mu_2 = V dp - S dT
\label{GD}
\end{equation}
to derive that $N_1(\partial\mu_1/\partial
p)_{Tx}+N_2(\partial\mu_2/\partial p)_{Tx}=V$. Combined with
$(\partial\mu_1/\partial p)_{Tx}=(\partial\mu_2/\partial p)_{Tx}$
(cf.~equation~\eqref{firstrelation}) this yields
\begin{equation}
  \ddc{\mu_1}{p}{Tx} =   \ddc{\mu_2}{p}{Tx} = \frac1n.
\label{secondrelation}
\end{equation}

We can write $c$ as a function of $x_1$ as $c=m_1 x_1/[m_1 x_1
+m_2(1-x_1)]$, which gives $d x_1/d c=\rho^2/(m_1 m_2 n^2)$. Using
also equation~\eqref{mudef}, we get for
equations~\eqref{firstrelation} and \eqref{secondrelation} in terms of
$c$ and $\mu$
\begin{eqnarray}
  \ddc{\mu}{c}{Tp} 
= \frac{dx_1}{dc}\ddc{\mu}{x_1}{Tp} 
=  \frac{\rho k_BT}{c(1-c)m_1 m_2 n}
\label{bmuciso}
\\
  \ddc{\mu}{p}{Tc} =  \ddc{\mu}{p}{Tx}  =
  \frac{m_2-m_1}{m_1 m_2 n}
\end{eqnarray}
From the fundamental relation for the Gibbs free energy $dG=-SdT+ V dp
+\mu_1 dN_1 +\mu_2 dN_2$ \cite[\S 8.7]{Reif}, one can deduce that
$(\partial\rho/\partial c)_{Tp}=-\rho^2(\partial\mu/\partial p)_{Tc}$,
whence
\begin{eqnarray}
  \ddc{\rho}{c}{Tp} =
\frac{(m_1-m_2)\rho^2}{m_1 m_2 n}.
\label{rhociso}
\end{eqnarray}

The quantities in equations~\eqref{bmuciso} and \eqref{rhociso} will
be needed in the treatment of the long time behaviour of the Green's
functions for mass transport in section~\ref{long-time-behavior}.
These are precisely of the form valid for ideal (``perfect'') mixtures
\cite[\S 92]{LandauLifshitzSP}, which is the reason why our treatment
is valid both for ideal mixtures and for isotopic binary mixtures.

Because in section~\ref{long-time-behavior} the results of the Green's
function approach for long times will be compared to those of
hydrodynamics, we will now review the appropriate hydrodynamic
equations and Green-Kubo formula for mass transport in ideal mixtures
and in an isotopic binary mixture. Since we restrict ourselves in this
paper to mass transport, the relevant conservation laws are only those
for the local densities $n_\lambda$, which follow from
equations~\eqref{1.1}, \eqref{1.8} and \eqref{1.11}:
\begin{equation}
  \dd{n_\lambda}{t} = -\bnabla \cdot \bi j^*_\lambda/m_\lambda
= -\bnabla \cdot (n_\lambda\bi u+\bi j_\lambda/m_\lambda)
\label{conservation}
\end{equation}
In terms of $\rho=m_1n_1+m_2n_2$ and $c=m_1n_1/\rho$, these equations
become \cite[ch.~2]{DeGrootMazur}
\begin{eqnarray}
 \frac{D\rho}{Dt}
  &=& -\rho\bnabla\cdot\bi u
\label{1}\\ 
  \rho\frac{Dc}{Dt}
  &=& -\bnabla\cdot\bi j_1
\label{4}
\end{eqnarray}
where we have used the material derivative $D/Dt=\partial/\partial
t+\bi u\cdot\bnabla$.  To make these equations into a closed set, $\bi
j_1$ needs to be expressed as a function of $\rho$ and $c$ (as well
as, in general, of $\bi u$ and $T$, which will be assumed to be
constant below).  Near equilibrium one can use linear laws as closure
relations.  Following a widely used choice of currents and forces in
the theory of irreversible thermodynamics \cite[ch.~3]{DeGrootMazur},
one has in general
\begin{eqnarray}
  \bi j_1 &=& -\frac{L_{11}}{T} [\bnabla\mu]_T  - \frac{L_{1 q}}{T}
 \bnabla \ln T
\label{once29}
\end{eqnarray}
Here the gradient $[\bnabla\mu]_T$ is taken at constant temperature
while the pressure $p$ and the concentration $c$ can vary. $L_{11}$
and $L_{1 q}$ are called phenomenological coefficients.

We will only consider here the case in which there is only a gradient
in the concentration $c$, while gradients in temperature $T$, pressure
$p$ and fluid velocity $\bi u$ are zero. In that case, one can write
equation~\eqref{once29} as
\begin{equation}
  \bi j_1 = - \frac{L_{11}}{T}\ddc{\mu}{c}{Tp}\bnabla c
\,=\, -\rho D \bnabla c
.
\label{J1stuf}
\end{equation}
Here the mutual diffusion constant $D = ({\partial\mu}/{\partial
c})_{Tp}L_{11}/(\rho T)$, or, with equation~\eqref{bmuciso} for binary
isotopic mixtures and ideal mixtures,
\begin{equation}
  D  =  \frac{k_BL_{11}}{c(1-c)m_1 m_2 n} .
\label{firstDM}
\end{equation}

The Green-Kubo formula for the phenomenological coefficient $L_{11}$
for mutual diffusion in a fluid is \cite{Wood89,Steele}:
\begin{eqnarray}
  L_{11}&=&\frac{1}{3k_BV}\int_0^\infty\!dt'\,
 \average{\tilde{\bi J}_1(0)\cdot\tilde{\bi
 J}_1(t')}
\label{L11GK}
\end{eqnarray}
where the average is over a grand canonical ensemble in the
thermodynamic limit and~$\tilde{\bi J}_\lambda$ is the microscopic
expression for the total current in the centre of mass frame given
by
\begin{eqnarray}
  \tilde{\bi J}_\lambda(t) &=& 
m_\lambda  \sum_{ i=1}^{N_\lambda}
  [\bi v_{\lambda  i}(t) 
-\bar{\bi v}]
\label{tildeJidef}
\end{eqnarray}
in which $\bar{\bi v}$ is the centre of mass velocity, i.e.
\begin{equation}
  \bar{\bi v}
  = 
(m_1 N_1+m_2 N_2)^{-1}
\sum_{\lambda =1}^{2}\sum_{ i=1}^{N_\lambda}m_\lambda\bi v_{\lambda  i}.
\label{vbar}
\end{equation}
The Green-Kubo formulae for $L_{1q}$ and other transport coefficients
are more complicated and not needed here; for these we refer to the
literature \cite{Wood89,Steele}.

Note that $\bar{\bi v}$ is not an ensemble average but an average over
particles taken per ensemble member, and that it is a constant of the
motion for that ensemble member.  With the inclusion of $\bar{\bi v}$
we have on a microscopic level the analogue of equation~\eqref{1.12}
i.e.\ $\tilde{\bi J}_1 + \tilde{\bi{J}}_2 = 0$.  This relation can be
used to find an alternative representation of $L_{11}$:
\begin{eqnarray}
  L_{11}
&=&-\frac{1}{3k_BV}\int_0^\infty\!dt'\,
 \average{\tilde{\bi J}_1(0)\cdot\tilde{\bi
 J}_2(t')}
\label{L11GKalt1}
\end{eqnarray}

Using equations~\eqref{tildeJidef} and \eqref{L11GKalt1} in
equation~\eqref{firstDM} for D, and that particles of the same
component are indistinguishable, we find that
\begin{eqnarray}
  D  &=& 
-\frac{1}{3c(1-c)nV} \int_0^\infty\!dt'\,
\average{
N_1 N_2 [\bi v_{1 1}(0)-\bar{\bi v}]
\cdot
[\bi v_{2 1}(t')-\bar{\bi v}]
}.
\label{DMGK}
\end{eqnarray}
In the thermodynamic limit ($N_1,N_2,V\to\infty$ keeping $N_1/V$ and
$N_2/V$ fixed and finite) we may factor the averages to leading order,
so that we get
\begin{equation}
  D  = - \frac{\rho^2}{m_1 m_2 n^2}f_{12},
\label{30prime}
\end{equation}
where \cite{McCallDouglass67}
\begin{eqnarray}
f_{12} &=& \frac13\int_0^\infty\!dt'\,
\average{
N
[\bi v_{11}(0)-\bar{\bi v}]
\cdot
[\bi v_{21}(t')-\bar{\bi v}]}.
\label{f12}
\end{eqnarray}
Similarly, for future use, one can also define, following
reference~\cite{McCallDouglass67}
\begin{eqnarray}
f_{\lambda\alpha\beta} &=&\frac13\int_0^\infty\!dt'\,
\average{N[\bi v_{\lambda1}(0)-\bar{\bi v}]\cdot[\bi
    v_{\lambda2}(t')-\bar{\bi v}]}
\label{fiab}
\\
D_\lambda
&=& \frac13\int_0^\infty\!dt'\,
\average{[\bi v_{\lambda1}(0)-\bar{\bi v}]\cdot[\bi
    v_{\lambda1}(t')-\bar{\bi v}]}.
\label{Di}
\end{eqnarray}
Note that the subscript $\alpha\beta$ does not denote indices but
instead signifies that correlations between different particles are
involved and that we added a factor $N$ in equations~\eqref{f12} and
\eqref{fiab} compared to the definition of $f_{12}$ and
$f_{\lambda\alpha\beta}$ in reference~\cite{McCallDouglass67} to get a
well-defined thermodynamic limit.  Due to the subtraction terms
$-\bar{\bi v}$ and the conservation of total momentum, one has the
relations\cite{McCallDouglass67}
\begin{eqnarray}
 f_{12} &=& 
- 
\frac{m_1}{m_2 x_2} (D_1 + x_1 f_{1\alpha\beta}) 
\label{relation1}
\\
f_{12} &=& 
-
 \frac{m_2}{m_1 x_1} (D_2 + x_2
 f_{2\alpha\beta}) 
\label{relation2}
\end{eqnarray}
Using these, one can rewrite the right hand side of
equation~\eqref{30prime} as
\cite{McCallDouglass67,DouglassFrisch69,ZhouMiller96}
\begin{eqnarray}
D  &=&
x_2 D_1
+
x_1 D_2
+ x_1 x_2
(f_{1\alpha\beta}+f_{2\alpha\beta}-2f_{12}).
\label{realDM}
\end{eqnarray}
The first part of this relation, i.e.\ the part involving only the
\emph{self} or \emph{intrinsic diffusion constants} $D_\lambda$, is
often used as if it were the full expression for $D$, which goes under
the name of the Darken \cite{Darken48} or Hartley-Crank equation
\cite{HartleyCrank49}. Douglass and co-workers were apparently the
first to have shown that corrections to the Darken-Hartley-Crank
equation involve the \emph{cross correlations} given by the
$f_{\lambda\alpha\beta}$ and $f_{12}$
\cite{McCallDouglass67,DouglassFrisch69}.

\section{Green's functions for mass transport in mixtures}
\label{greens-funct-meth}

\subsection{General formulation}

The hydrodynamic treatment only works near (local) equilibrium. For
other situations i.e.\ on small length and time scales, a new
description is necessary. For self-diffusion the short time behaviour,
as well as the long time behaviour, could be successfully described by
using Green's functions \cite{Kincaid95}.  A similar description for
heat transport gives good numerical agreement with simulations as well
\cite{KincaidCohen02a,KincaidCohen02b,CohenKincaid02}.

Analogous to the Green's function approach for self-diffusion we want
to write the $\bi r$ and $t$ dependent densities $n_\lambda$ here as
integrals over Green's functions that give the probability for a
representative particle of the appropriate component to be at position
$\bi r$ at time $t$ given that it was at position $\bi r'$ at time
$0$. Starting from equation~\eqref{1.1} we can write ($\lambda=1,2$)
\begin{eqnarray}
\label{2.1}
  n_\lambda(\bi r, t) 
  &=& 
  N_\lambda  
  \average{\delta\mathbf(\bi r-\bi r_{\lambda1}(t)\mathbf)}
  _{\mathrm{ne}}
  =
  \int_V d\bi r'\, G_\lambda(\bi r,\bi r',t)
  \,n_\lambda(\bi r',0)  
  ,
\end{eqnarray}
where the Green's function $G_\lambda$ is defined as
\begin{eqnarray}
  G_\lambda(\bi r,\bi r',t) &=& 
\average{\delta\mathbf(\bi r-\bi r'-\Delta\bi r_{\lambda1}(t)\mathbf)}_{\lambda\bi r'}
.
\label{2.2}
\end{eqnarray} 
Here $\Delta\bi r_{\lambda1}(t)=\bi r_{\lambda1}(t) -\bi
r_{\lambda1}(0)$ is the displacement of particle $1$ of component
$\lambda$ while $\average{}_{\lambda\bi r'}$ is the average over
initial conditions for which $\bi r_{\lambda1}(0)=\bi r'\!$.  For an
arbitrary phase function $A(\Gamma)$, this conditional average is
defined as
\begin{eqnarray}
\average{A(\Gamma)}_{\lambda\bi r'}
&=&
\frac{
  \average{A(\Gamma)\,\delta\mathbf(\bi r'-\bi
    r_{\lambda1}(0)\mathbf)}_{\mathrm{ne}}
}
{
 \average{ \delta\mathbf(\bi r'-\bi r_{\lambda1}(0)\mathbf)}_{\mathrm{ne}}
}.
\label{2.4}
\end{eqnarray}

We now make two remarks about this definition of the Green's functions
and the conditional average.  First, for initial conditions where
$n_\lambda(\bi r',0)=N_\lambda\average{ \delta\mathbf(\bi r'-\bi
r_{\lambda1}(0)\mathbf)}_{\mathrm{ne}}=0$ for some $\bi r'$, the
right-hand side of equation~\eqref{2.4} seems ill defined. The way we
choose to define it for such $\bi r'$ will not matter since
$\average{\delta(\bi r-\bi r'-\Delta\bi r_{\lambda1}(t))}_{\lambda\bi
r'}$ is multiplied by $n_\lambda(\bi r',0)=0$ in equation~\eqref{2.1}
anyway.\footnote{However, to have a proper definition everywhere, one
could follow the convention that $\average{A}_{\lambda\bi
r'}=\average{A}_{\mathrm{ne}}$ for values of $\bi r'$ where
$n_\lambda(\bi r',0)=0$. An alternative would be to introduce a
``tracer particle'' of component $\lambda$ at $\bi r'$ but this works
only in the thermodynamic limit.}  Second, equation~\eqref{2.4}
differs in normalization from the definition of $\average{\,}_{r'}$ in
the heat pulse case of
references~\cite{KincaidCohen02a,KincaidCohen02b,CohenKincaid02}, in
which the conditional averages were only properly normalized when the
system is initially of uniform density (which was the case in
references~\cite{KincaidCohen02a,KincaidCohen02b,CohenKincaid02}).  In
contrast, the definition of the conditional average in
equation~\eqref{2.4} is always normalized i.e.\
$\average{1}_{\lambda\bi r'}=1$. This definition is also more
analogous to the self-diffusion case \cite{Kincaid95} because of the
appearance of $n_\lambda(\bi r',0)$ inside the integrals in
equation~\eqref{2.1} instead of a single factor of the average density
$N_\lambda/V$ in front of the integral.

\subsection{Expansion around Gaussian behaviour}

Next the Green's functions $G_\lambda$ are Fourier transformed with
respect to $\bi r - \bi r'$, i.e.\ one considers
\begin{eqnarray}
\label{2.5}
F_\lambda(\bi k,\bi r', t) 
&=& \int_V d\bi r\,\mathrm e^{\mathrm i\bi k\cdot(\bi r-\bi r') }
G_\lambda(\bi r,\bi r',t)
\end{eqnarray}
Using equation~\eqref{2.2} and expanding in $\bi k$, this gives
\begin{eqnarray}
  F_\lambda(\bi k,\bi r', t) &=& \average{
                \exp[{\mathrm i\bi k\cdot \Delta\bi r_{\lambda1}(t)}]
                }_{\lambda\bi r'}
\:=\:
\sum_{n=0}^{\infty} 
\frac{(\mathrm i k)^n}{n!}
\mu_{\lambda n}(\hat{\bi k}, \bi r', t)
\label{2.6}
\end{eqnarray}
where $k=|\bi k|$, $\hat k=\bi k/k$ and the moments are defined as
\begin{equation}
\mu_{\lambda n}(\hat{\bi k},\bi r', t)=
 \bigaverage{\left[\hat{\bi k}\cdot\Delta\bi r_{\lambda1}(t)\right]^n}_{\lambda\bi r'}.
\label{momdef}
\end{equation}
Since divergences would occur if one would directly take the inverse
Fourier transform term by term in equation~\eqref{2.6}, an alternative
expansion will be used. For given direction $\hat{\bi k}$, $F_\lambda$
is the generating function of the cumulants of the displacement of the
particle in that direction, \cite{Cramer46,VanKampen,VanZonCohen05c}
so that one can write
\begin{equation}
    F_\lambda(\bi k,\bi r', t)
    =
    \exp \sum_{n=1}^\infty \frac{(ik)^n}{n!} \kappa_{\lambda
    n}(\hat{\bi k}, \bi r', t)
\label{sstarr}
\end{equation}
The cumulants $\kappa_{\lambda n}$ and $\mu_{\lambda n}$ are related
through \cite{VanZonCohen05c}
\begin{equation}
  \kappa_{n\lambda} \:=\: -\,n!
\mathop{\sum_{\{p_\ell\geq 0\}}}_{\sum_{\ell=1}^\infty \ell p_\ell = n}
\Big(\sum_{\ell=1}^\infty p_\ell-1\Big)!
\prod_{\ell=1}^\infty \frac{\big[
-\mu_{\lambda \ell} / \ell!
\big]^{p_\ell}
}{p_\ell!}
\label{kappaintermsofmu}
\end{equation}
where we dropped the arguments $\hat{\bi k}$, $\bi r'$ and $t$ of
$\mu_{\lambda\ell}$ and $\kappa_{\lambda n}$, as we will do below as
well.  For instance, for the first few $\kappa_{\lambda n}$,
equation~\eqref{kappaintermsofmu} becomes
\begin{eqnarray}
\kappa_{\lambda 1}
&=&
\mu_{\lambda 1}
\\
\kappa_{\lambda2}
&=&
\mu_{\lambda 2}-\mu_{\lambda 1}^2
\\
\kappa_{\lambda 3}
&=&\mu_{\lambda 3}-3\mu_{\lambda 1}\mu_{\lambda 2}+2\mu_{\lambda 1}^3
\\
\kappa_{\lambda 4}
&=&\mu_{\lambda 4}-4\mu_{\lambda 1}\mu_{\lambda 4}
-3\mu_{\lambda2}^2
+12 \mu_{\lambda2}\mu_{\lambda1}^2
-6\mu_{\lambda1}^4.
\end{eqnarray}

The Fourier transform in equation~\eqref{sstarr} is first rearranged
by factoring out a Gaussian expression
\cite{Kincaid95,KincaidCohen02a,KincaidCohen02b,CohenKincaid02}.  This
can be done in several ways, but here we write
\begin{eqnarray}\label{resum2}
  F_\lambda(\bi k,\bi r', t) &=& 
f_\lambda(\bi k, \bi r', t) 
\left[1 + \sum_{n=3}^\infty 
b_{\lambda n} (\mathrm i k)^{  n}\right].
\end{eqnarray}
and we demand that the first correction term to the Gaussian
$f_\lambda(\bi k, \bi r', t)$ in equation~\eqref{resum2} is only
$\mathcal O(k^3)$, so that
\begin{eqnarray}
  f_\lambda(\bi k, \bi r', t)  &=& 
\exp\left[
\mathrm i \kappa_{\lambda1} k -\case12
\kappa_{\lambda2} k^2
\right]
\label{Gkdef}
\end{eqnarray}
with $\kappa_{\lambda1}$ the mean of the displacement and
$\kappa_{\lambda2}$ its variance, explicitly given by
\begin{eqnarray}
\label{msdef}
\kappa_{\lambda1}&=& 
\average{\hat{\bi{k}}\cdot\Delta\bi r_{\lambda1}(t)}_{\lambda\bi r'}
\\
\label{Vsdef}
 \kappa_{\lambda2} &=& 
\average{
[\hat{\bi{k}}\cdot\Delta\bi r_{\lambda1}(t)-\kappa_{\lambda1}(\hat{\bi k}, \bi r', t)]^{ 2}}_{\lambda\bi r'}.
\end{eqnarray}
The coefficients $b_{\lambda n}$ in equation~\eqref{resum2} can be
determined by gathering terms of like order in $k$ in the
$k$-expansion of $F_\lambda(\bi k, \bi r',t)/f_\lambda(\bi k, \bi
r',t)$.  This gives
\begin{equation}
  b_{\lambda n} \:=\:
\mathop{\sum_{\{p_\ell\geq 0\}}}_{\sum_{\ell=3}^\infty \ell p_\ell = n}
\prod_{\ell=3}^\infty\:\:
\left[
\frac{1}{p_\ell!}\left(\frac{\kappa_{\lambda \ell}}{\ell!}\right)^{p_\ell}
\right]
.
\label{bsdef}
\end{equation}
where the arguments $\hat{\bi k}$, $\bi r'$ and $t$ of $b_{\lambda n}$
and $\kappa_{\lambda n}$ have been suppressed.  The first few
coefficients $b_{\lambda n}$ are given by
\begin{eqnarray}
\label{bn3}
b_{\lambda 3}&=&\kappa_{\lambda 3}/3!
\\ 
b_{\lambda 4}&=&\kappa_{\lambda 4}/4!
\label{bn4}
\end{eqnarray}
Up to a constant prefactor, the coefficients $\kappa_{\lambda1}$
(mean), $\kappa_{\lambda2}$ (variance), $b_{\lambda 3}$ (skewness),
$b_{\lambda 4}$ (kurtosis) and $b_{\lambda 5}$ are pure cumulants,
however starting from $n=6$ the coefficients $b_{\lambda n}$ can only
be written as combinations of cumulants, e.g.\ $b_{\lambda 6}=
(\kappa_{\lambda 6}+10\kappa_{\lambda 3}^2)/6!$.

Term by term the expression in equation~\eqref{resum2} can be Fourier
inverted.  However, one has to keep the $\hat{\bi k}$ dependence of
the $\kappa_{\lambda n}$ in mind. For this $\hat{\bi k}$ dependence,
we can write, using equations~\eqref{momdef}, \eqref{kappaintermsofmu}
and \eqref{bsdef},
\begin{eqnarray}
  \mu_{\lambda n} &=& {\hat{\bi k}}^n : \mathsf M_{\lambda n}
\label{muintermsofM}
\\
  \kappa_{\lambda n} &=& {\hat{\bi k}}^n : \mathsf K_{\lambda n}
\label{kappaintermsofK}
\\
  b_{\lambda n} &=& {\hat{\bi k}}^n : \mathsf B_{\lambda n}
\label{bintermsofB}
\end{eqnarray}
where $\mathsf M_{\lambda n}$, $\mathsf K_{\lambda n}$ and $\mathsf
B_{\lambda n}$ are $\bi r'$ and $t$ dependent tensors of rank $n$.
Here a tensor notation is used. In this notation, ``$:$'' denotes a
scalar product (summation over all $n$ indices), while powers of
tensors are $\mathsf A^{ n}=\mathsf A\otimes\mathsf
A\cdots\otimes\mathsf A$ and yield symmetric tensors of rank $n$.  The
tensors $\mathsf M_{\lambda n}$, $\mathsf K_{\lambda n}$ and $\mathsf
B_{\lambda n}$ are chosen to be symmetric, and follow from
equations~\eqref{momdef}, \eqref{kappaintermsofmu}, \eqref{bsdef} and
equations~\eqref{muintermsofM}-\eqref{bintermsofB}:\footnote{\label{symprod}To
avoid having to use explicit indices, we use here a symmetrized tensor
product, such that $\mathsf A\mathsf B=[\mathsf A\otimes \mathsf
B]_S$, where $[\,]_{\rm S}$ denotes the symmetric part of a tensor
e.g.\ $[\mathsf A]_{{\rm S}ij}=(\mathsf A_{\lambda j}+\mathsf
A_{ji})/2!$ for a rank two tensor, $[\mathsf A]_{{\rm S}ijk}=(\mathsf
A_{\lambda jk}+\mathsf A_{jik}+\mathsf A_{\lambda kj}+\mathsf
A_{kji}+\mathsf A_{kij}+\mathsf A_{jki})/3!$ for rank three, etc.  }
\begin{eqnarray}
  \mathsf M_{\lambda n} &=& 
\average{[\Delta\bi r_{\lambda1}(t)]^n}_{\lambda\bi r'}
\label{Mdef}
\\
  \mathsf K_{\lambda n} &=& -\,n!
\mathop{\sum_{\{p_\ell\geq 0\}}}_{\sum_{\ell=1}^\infty \ell p_\ell = n}
\Big(\sum_{\ell=1}^\infty p_\ell-1\Big)!
\prod_{\ell=1}^\infty \frac{\big[
-\mathsf M_{\lambda \ell}/\ell!
\big]^{p_\ell}
}{p_\ell!}
\label{KintermsofM}
\\
  \mathsf B_{\lambda n} &=&
\mathop{\sum_{\{p_\ell\geq 0\}}}_{\sum_{\ell=3}^\infty \ell p_\ell = n}
\prod_{\ell=3}^\infty\:\:
\left[
\frac{1}{p_\ell!}\left(\frac{\mathsf K_{\lambda \ell}}{\ell!}\right)^{p_\ell}
\right]
.
\label{BintermsofK}
\end{eqnarray}
Using these definitions, the Fourier inverse of $f_\lambda$ of
equation~\eqref{Gkdef} is
\begin{equation}
 g_\lambda(\bi r, \bi r', t) 
=
\frac{\exp[-\case12 
(\bi r-\bi r'-\mathsf K_{\lambda1})\cdot
\mathsf K_{\lambda2}^{-1}\cdot(\bi r-\bi r'-\mathsf K_{\lambda1})]}
{\sqrt{\det(2\pi\mathsf K_{\lambda2})}}.
\label{gdef2}
\end{equation}
where
\begin{eqnarray}
\label{K1def}
\mathsf K_{\lambda1}&=& \average{\Delta\bi r_{\lambda1}(t)}_{\lambda\bi r'}
\\
\label{K2def}
 \mathsf K_{\lambda2} &=& 
\average{
[\Delta\bi r_{\lambda1}(t)-\mathsf K_{\lambda1}]^{ 2}}_{\lambda\bi r'}.
\end{eqnarray}
Using equations~\eqref{resum2} and \eqref{bintermsofB} and that the
Fourier inverse of $F_\lambda$ is $G_\lambda$ and that of $f_\lambda$
is $g_\lambda$, we obtain
\begin{eqnarray}
   G_\lambda(\bi r,\bi r',t) &=& \left[
     1 + \sum_{n=3}^\infty\mathsf B_{\lambda n}: (-\bnabla)^{  n}
\right] g_\lambda(\bi r, \bi r', t)
\label{Gspatorig}
\\
&=&
 \left[
     1 + \sum_{n=3}^\infty\mathsf B_{\lambda n}: \mathsf H_n\Big(\bi
     r-\bi r'-\mathsf K_{\lambda1}\,,\,\case12\mathsf K_{\lambda2}^{-1}\Big)
\right] 
g_\lambda(\bi r, \bi r', t)
.
\label{Gspat}
\end{eqnarray}
Here the $\mathsf H_{n}$ are polynomial tensors in $\bi s=\bi r-\bi
r'-\mathsf K_{\lambda1}$ given (with $\mathsf A= \case12\mathsf
K_{\lambda2}^{-1}$) by
\begin{equation}
\label{tensHndef}
  \mathsf H_n(\bi s, \mathsf A) = \sum_{m=0}^{\lfloor\frac{n}{2}\rfloor}
 \frac{(-1)^mn!}{m!(n-2m)!}
\mathsf A^{  m} (2\mathsf A\cdot\bi s)^{  n-2m}
.
\end{equation}
where $\lfloor\frac{n}{2}\rfloor$ is the largest integer number less
than or equal to $\frac{n}{2}$.  Explicitly, the first few polynomials
$\mathsf H_{n\geq3}(\bi s,\mathsf A)$ are given by
\begin{eqnarray}
\label{Hn3}
\mathsf H_3(\bi s,\mathsf A) 
&=& 
4[2(\mathsf A\cdot\bi s)^{  3}
  -3\mathsf A (\mathsf A\cdot\bi s)]
\\ 
\label{Hn4}
\mathsf H_4(\bi s,\mathsf A) &=&
4[4 (\mathsf A\cdot\bi s)^{  4}-12
\mathsf A (\mathsf A\cdot\bi s)^{  2}
+3
\mathsf A^2
].
\end{eqnarray}
The coefficients in the expressions of the right hand sides for these
$\mathsf H_n$ are precisely those of the Hermite polynomials. They can
in fact be related to the Hermite polynomials, in that the $n$th
Hermite polynomial can be expressed as $H_n(s_x)=\mathsf H_n(\bi s,
\mathbbm{1}):\hat{\bi x}^n$ (using Gradshteyn and Ryzhik 8.950.2
\cite{GradshteynRyzhik} with $\bi s=(s_x,s_y,s_z)$, $\mathbbm{1}$ the
identity matrix, $\mathsf A:\hat{\bi x}^n$ the (all) $x$ component of
$\mathsf A$, and noting that $2^m (2m-1)!!/(2m)! = 1/m!$).

\subsection{Reduction to one-dimensional Green's functions}

The somewhat cumbersome tensorial notation of the last section can be
avoided in cases where the initial distribution $\mathcal
P_{\mathrm{ne}}(\Gamma)$ is only nonuniform in the $x$ direction and
translation invariant in the $y$ and $z$ directions, such as in
previous work on the Green's functions
\cite{Kincaid95,KincaidCohen02a,KincaidCohen02b,CohenKincaid02}.  In
such a situation one can use the reduced Green's function
\begin{eqnarray}
  G^x_\lambda(x, x',t) &\equiv& 
 \int\! dy'\!\int\! dz'\:G_\lambda(\bi r,\bi r',t)
=
\average{\delta(x-x'-\Delta x_{\lambda1}(t)\mathbf)}_{\lambda x'}
.
\label{2.2red}
\end{eqnarray} 
where $\Delta x_{\lambda1}(t)=x_{\lambda1}(t) -x_{\lambda1}(0)$ is the
displacement of particle $1$ of component $\lambda$ in the $x$
direction.  Note that the $y$ and $z$ dependence has dropped out
because of the translation invariance of $\mathcal
P_{\mathrm{ne}}(\Gamma)$ in those directions and the translation
invariance of the dynamics i.e.\ of the Hamiltonian.  In terms of the
reduced Green's function equation~\eqref{2.1} becomes
\begin{eqnarray}
\label{2.1red}
  n_\lambda(x, t) &=& 
\int dx'\: G^x_\lambda(x,x',t) \,n_\lambda(x',0).
\end{eqnarray}
To see how the reduction works out for expansion \eqref{Gspat}, it is
easier to start from the original expansion in
equation~\eqref{Gspatorig}. Noting that because of translation
invariance the $\mathsf K_{\lambda1}(\bi r',t)$ and consequently the
$\mathsf B_{\lambda n}(\bi r',t)$ do not depend on $y'$ or $z'$, we
can integrate equation~\eqref{Gspatorig} over $y'$ and $z'$ to find
\begin{eqnarray}
   G^x_\lambda(x,x',t) &=& \left[
     1 + \sum_{n=3}^\infty b^x_{\lambda n}
\left(-\dd{}{x}\right)^n
\right] g^x_\lambda(x, x', t)
\label{letssee}
\end{eqnarray}
where $b^x_{\lambda n}=\mathsf B_{\lambda n}:\hat{\bi x}^n$ and
\begin{eqnarray}
  g^x_\lambda(x, x',t) &=& 
 \int\! dy'\!\int\! dz'\:g_\lambda(\bi r,\bi r',t)
= \frac{\displaystyle\exp\left[-\frac{(x-x'-\kappa^x_{\lambda1})^2}{2\kappa_{\lambda2}^{x}}\right]}
 {\sqrt{2\pi \kappa_{\lambda2}^{x}}},
\label{defgi}
\end{eqnarray}
where $\kappa^x_{\lambda n}=\mathsf K_{\lambda n}:\hat{\bi x}^n$ and
we used that the $xy$ and $xz$ components of the matrix $\mathsf
K_{\lambda2}$ are zero by translation symmetry in the $y$ and $z$
direction (which also pertains to the inverse of this matrix).
Working out the derivatives to $x$ in equation~\eqref{letssee} gives
\begin{eqnarray}
   G^x_\lambda(x,x',t) &=& \Bigg[
     1 + \sum_{n=3}^\infty \frac{b^x_{\lambda n}}
     {(2\kappa^{x}_{\lambda2})^{n/2}}
     H_n\Bigg(\frac{x-x'-\kappa^x_{\lambda1}}
     {\sqrt{2\kappa^{x}_{\lambda2}}}\Bigg)
\Bigg] g^x_\lambda(x, x', t),
\label{onedim}
\end{eqnarray}
where the $H_n$ are true Hermite polynomials.

We note that to obtain the $b^x_{\lambda n}$ and $\kappa^x_{\lambda
n}$ it is possible to avoid the tensors through which they have been
defined, since by comparing the definitions $\kappa^x_{\lambda
n}=\mathsf K_{\lambda n}:\hat{\bi x}^n$ and $b^x_{\lambda n}=\mathsf
B_{\lambda n}:\hat{\bi x}^n$ with the expressions in
equation~\eqref{kappaintermsofK} and equation~\eqref{bintermsofB}, one
sees immediately that one only needs to set $\hat{\bi k}$ equal to
$\hat{\bi x}$ in the expressions for $b_{\lambda n}$ and
$\kappa_{\lambda n}$ to obtain $b^x_{\lambda n}$ and $\kappa_{\lambda
n}^x$.

We note that the resummation of $G_\lambda$ leading to
equations~\eqref{Gspat} and \eqref{onedim} is somewhat different from
that used for the non-equilibrium heat transport case of
references~\cite{KincaidCohen02a,KincaidCohen02b,CohenKincaid02} where
separate Gaussian factors were introduced for the even terms and the
odd terms in $k$ (i.e.\ even and odd $n$), respectively, rather than a
single overall Gaussian factor as we used in
equation~\eqref{resum2}. A result was then found in terms of Sonine
polynomials $S^{(n)}_m(x)$ \cite{ChapmanCowling}. The present
formalism is nonetheless very similar since the Hermite polynomials
are special cases of the Sonine (or Laguerre) polynomials (i.e.\
$H_{2n}(x)=(-1)^n2^{2n}n!S_{-1/2}^{(n)}(x^2)$ and
$H_{2n+1}(x)=(-1)^n2^{2n+1}n!xS_{1/2}^{(n)}(x^2)$, see references
\cite[p.~127]{ChapmanCowling} and \cite[8.972]{GradshteynRyzhik}).
Leaving out the odd terms in equation~\eqref{onedim} furthermore
yields the exact same form as that in the self-diffusion result in
equation (6) of reference~\cite{Kincaid95} (with $\rho_1=\case12
\kappa_{\lambda2}^{x}$, $\rho_{2}=b^x_{\lambda 4}$ and
$\rho_{3}=b^x_{\lambda 6}$).

\subsection{Practical considerations}

Substituting the expression for $G_\lambda$ in equation~\eqref{Gspat}
into expression \eqref{2.1} (or equation~\eqref{onedim} into
equation~\eqref{2.1red}) for the local densities $n_\lambda$ in terms
of the Green's functions, leads to an expression for those densities
involving an infinite sum over $n$.  To make the Green's function
approach practical the dynamical averages occurring in $\mathsf
M_{\lambda n}$ in equation~\eqref{Mdef} (and in $\mathsf K_{\lambda
n}$ and $\mathsf B_{\lambda n}$ through equations~\eqref{KintermsofM}
and \eqref{BintermsofK}) need to be calculated, for instance by
molecular dynamics. Since one can calculate only finitely many
quantities in practise the infinite sum in equation~\eqref{Gspat} has
to be truncated at some point.

Having chosen the Gaussian factor $g_\lambda$ in
equation~\eqref{resum2} such that corrections are at most $\mathcal
O(k^3)$, the Gaussian approximation
\begin{equation}
\label{secondapprox}
  G_\lambda(\bi r,\bi r',t) \approx g_\lambda(\bi r,\bi
  r',t)
=
\frac{\exp[-\case12 
(\bi r-\bi r'-\mathsf K_{\lambda1})\cdot
\mathsf K_{\lambda2}^{-1}\cdot(\bi r-\bi r'-\mathsf K_{\lambda1})]}
{\sqrt{\det(2\pi\mathsf K_{\lambda2})}}.
\end{equation}
(cf.~equation~\eqref{gdef2}) is expected to describe the correct long
time limiting behaviour, similar to the self-diffusion case
\cite{Kincaid95}.  To confirm this expectation we will look in detail
at the behaviour in the long time limit in
section~\ref{long-time-behavior}.  For intermediate
(non-infinitesimally small) times, it will be necessary to include
some of the correction terms proportional to $\mathsf B_{\lambda n}$
in equation~\eqref{Gspat} e.g.\ (using
equations~\eqref{Mdef}--\eqref{BintermsofK}, \eqref{Hn3} and
\eqref{Hn4})
\begin{eqnarray}
G_\lambda(\bi r,\bi r',t) &\approx\:  g_\lambda(\bi r,\bi r',t)
\bigg[&
1 + \mathsf K_{\lambda 3} :(\bi q_\lambda^{  3}-3\mathsf K_{\lambda2}^{-1} \bi q_\lambda)
\nonumber\\&&
+\mathsf K_{\lambda 4}
:( \bi q_\lambda^{  4}-6\mathsf K_{\lambda2}^{-1} \bi q_\lambda^{  2}
  +3[\mathsf K_{\lambda2}^{-1}]^{  2})
+\:\ldots\,
\bigg],
\label{somecorrections}
\end{eqnarray}
where $\bi q_\lambda=\mathsf K_{\lambda2}^{-1}\cdot(\bi r-\bi
r'-\mathsf K_{\lambda1})$. Checking whether these first few terms for
the Green's functions describe the intermediate time behaviour\
accurately will be done in future numerical work.

\subsection{Connection with the Van Hove self-correlation function}

For equilibrium initial conditions (i.e.\ replacing
$\average{}_{\lambda\bi r'}$ by $\average{}\,$), the Green's function
$G_\lambda(\bi r,\bi r',t)$ depends only on the difference $|\bi r-\bi
r'|$ and reduces to the Van Hove self-correlation function
\cite{VanHove54,BoonYip80}.  Indeed, the self-part $G_s(\bi r-\bi
r',t)$ of the Van Hove function also has the physical interpretation
of the probability for a particle to be at position $\bi r$ given that
it was at position $\bi r'$ a time $t$ earlier. Thus the Green's
function $G_\lambda(\bi r,\bi r',t)$ can be seen as a nonequilibrium
extension of $G_s(\bi r-\bi r',t)$. It is at present not clear whether
this means that Green's functions could be observed in incoherent
neutron scattering experiments like the Van Hove self-correlation
function can \cite{VanHove54}.

Similar expansions as in equation~\eqref{resum2} have been used for
the Van Hove self-correlation function $G_s$ in a single component
fluid in equilibrium
\cite{Rahmanetal62,Rahman64,NijboerRahman66,BoonYip80,VanZonCohen05c},
where odd terms in $\bi k$ are absent.  The Gaussian approximation to
the Van Hove self-correlation function turns out a good description in
the very short and the long time limits. Non-Gaussian corrections are
important for intermediate times, where including just the first
correction term (involving the fourth moment) captures most of the
physical behaviour \cite{Rahmanetal62,Rahman64,BoonYip80}.  For more
information, we refer to chapter 4 of reference~\cite{BoonYip80} which
contains a nice overview of different approaches to $G_s$.

\section{Limiting behaviour of the Green's functions at long times}
\label{long-time-behavior}

As a first approximation in investigating the large time and length
scale behaviour of the Green's functions, we realize that at long
times gradients get small so that $\mathcal O(k^3)$ terms become
unimportant and the Gaussian approximation, found by neglecting the
terms proportional to $\mathsf B_{\lambda n}$ in
equation~\eqref{Gspat} ($\mathcal O(k^3)$ terms in
equation~\eqref{resum2}) should capture all of the relevant dynamics
for long times, since the hydrodynamic equations also contain at most
second orders in gradients. This will now be checked.

Taking into account only the Gaussian term in the resummation in
equation~\eqref{Gspat}, we get the Gaussian approximation in
\eqref{secondapprox}, with $\mathsf K_{\lambda1}$ and $\mathsf
K_{\lambda2}$ given in equations~\eqref{K1def} and \eqref{K2def}.  We
emphasize that the present approach differs from the self-diffusion
case \cite{Kincaid95} where no odd terms in $\bi k$ existed. But in
the mutual diffusion case odd terms such as ${\mathsf
K}_{\lambda1}(\bi r',t)$ cannot be neglected. These terms are of the
form $\average{\Delta \bi r_{\lambda1}(t)}_{\lambda\bi r'}$ so they
represents net particle displacement or drift. In a diffusing system
particles do move in a preferred direction, namely opposite to the
gradient, so these terms will not be zero and will in fact be
sensitive to the gradients.

Using the Gaussian approximation $G_\lambda\approx g_\lambda$ in
equation~\eqref{secondapprox}, a diffusion-like equation can be
derived for the Green's function $G_\lambda$. The easiest way to
derive this is by taking the time derivative on both sides of
equation~\eqref{Gkdef}, thus obtaining an equation for the Fourier
transform $F_\lambda\approx f_\lambda$, and then taking the inverse
Fourier transform to get the corresponding equation for
$G_\lambda\approx g_\lambda$. This leads to
\begin{eqnarray}
\label{diffdrift}
    \dd{}{t}
 G_\lambda (\bi r,\bi r',t) 
\approx -\dot{\mathsf K}_{\lambda1}\cdot \bnabla G_\lambda (\bi r,\bi r',t)
    + \case12\bnabla\cdot\dot{\mathsf K}_{\lambda2} \cdot
\bnabla  G_\lambda (\bi r,\bi r',t)
\end{eqnarray}
where $\dot{\mathsf K}_{\lambda1}=\partial \mathsf
K_{\lambda1}/\partial t$ and $\dot{\mathsf K}_{\lambda2}=\partial
\mathsf K_{\lambda2}/\partial t$.  Using equation~\eqref{2.1}, which
relates $G_\lambda$ to $n_\lambda$, this implies that
\begin{eqnarray}
\label{start2}
    \dd{n_\lambda}{t} &\approx &
-\bnabla \cdot\Bigg[
 \int_V d\bi r'\,
\dot{\mathsf K}_{\lambda1}(\bi r',t) G_\lambda(\bi r,\bi r',t) n_\lambda(\bi r',0)
\nonumber\\&&\qquad
+\int_V d\bi r'\,
\case12\dot{\mathsf K}_{\lambda2}(\bi r',t)\cdot\bnabla G_\lambda(\bi r,\bi r',t) n_\lambda(\bi r',0)
\Bigg].
\end{eqnarray}

In equation~\eqref{start2} we focus first on $\dot{\mathsf
K}_{\lambda2}$. From equation~\eqref{K2def} we have
\begin{eqnarray}
\label{nnn}
  \dot{\mathsf K}_{\lambda2}(\bi r', t) &=& 
\dd{}{t}\average{
 [
 \Delta\bi r_{\lambda1}(t)
 -\average{
   \Delta\bi r_{\lambda1}(t)
 }_{\lambda\bi r'}
 ]^{  2}
}_{\lambda\bi r'}
\nonumber\\
&=& 
2\average{
  [
  \bi v_{\lambda1}(t)
 -\average{\bi{v}_{\lambda1}(t)}_{\lambda\bi r'}
 ] [
 \Delta\bi r_{\lambda1}(t)
 -\average{\Delta\bi r_{\lambda1}(t)}_{\lambda\bi r'}
 ]
}_{\lambda\bi r'}
\nonumber\\
& =&
2\int_0^t dt'\,\average{
  [
  \bi v_{\lambda1}(t)
 -\average{\bi{v}_{\lambda1}(t)}_{\lambda\bi r'}
 ] [
  \bi v_{\lambda1}(t')
 -\average{\bi{v}_{\lambda1}(t')}_{\lambda\bi r'}
 ]
}_{\lambda\bi r'}
\end{eqnarray}
(using the symmetrized tensor product explained in the footnote on
page~\pageref{symprod}). Equation \eqref{nnn} looks surprisingly like
a Green-Kubo formula for the self-diffusion coefficients except for
the conditional average $\average{}_{\lambda\bi r'}$ and the
integration to $t$. It seems reasonable to assume that for long times
$t$ the condition that a particle was at some position $\bi r'$ at
time $0$ should have a negligible effect on how fast it is diffusing
at time $t$. In other words it seems reasonable that the limit of
$t\to\infty$ of the above expression is independent of~$\bi r'$:
\begin{eqnarray}
\lim_{t\to\infty}\dot{\mathsf K}_{\lambda2}(\bi r',t) &=& 
\lim_{t\to\infty} 2 \int_0^t dt'\,
\average{
[
\bi  v_{\lambda1}(t)-\average{\bi v_{\lambda1}(t)}
] [
\bi v_{\lambda1}(t')
-\average{\bi v_{\lambda1}(t')}
]
}
\end{eqnarray}
For $t\to\infty$ the (closed) system will have reached equilibrium, so
we can use the time translation invariance and isotropy of the
equilibrium distribution and that $\average{\bi v_{\lambda1}(t)}=0$
for all $t$, to write
\begin{eqnarray}
\lim_{t\to\infty} \dot{\mathsf K}_{\lambda2}(\bi r',t) &=& 
  \frac23\int_0^\infty dt'\,\average{\bi v_{\lambda1}(0)\cdot
\bi  v_{\lambda1}(t')} \mathbbm{1}
\:=\: 2 D_\lambda \mathbbm{1}
,
\label{VtoD}
\end{eqnarray}
where we used the definition of $D_\lambda$ in equation~\eqref{Di}. (Note that
formally one needs to take the thermodynamic limit before taking
$t\to\infty$.)

Assuming now that $\dot{\mathsf K}_{\lambda2}(\bi r', t)$ attains its
limiting value $2 D_\lambda$ fast enough, we can replace $\dot{\mathsf
K}_{\lambda2}(\bi r', t)$ by that value in equation~\eqref{start2} for
long times.  It is then independent of $\bi r'$ and can be taken
outside of the integral in equation~\eqref{start2}. (Another way to
see that the $\bi r'$ dependence of $\dot{\mathsf K}_{\lambda2}$
should drop out is by noting that this $\bi r'$ dependence will just
give additional gradients in $\bi r$ if one expands around $\bi r$,
but the $\dot{\mathsf K}_{\lambda2}$ term in equation~\eqref{start2}
is already of the highest order in gradients required for
hydrodynamics, i.e.\ second order in the gradients.)  Equation
\eqref{start2} then becomes
\begin{eqnarray}
    \dd{n_\lambda}{t} &\approx &
-\bnabla \cdot
\int_V d\bi r'\,
\dot{\mathsf K}_{\lambda1}(\bi r', t) G_\lambda(\bi r,\bi r',t)
n_\lambda(\bi r',0)
\,+ \,D_\lambda\bnabla^2 n_\lambda.
\label{complex}
\end{eqnarray}

Because the first expression on the right-hand side of
equation~\eqref{complex} cannot be further calculated without
additional assumptions, to see whether it describes long time
behaviour in accordance with hydrodynamics, one next makes the
essential assumption of local equilibrium (which is also used to
derive hydrodynamics in other ways).

To keep the derivation of hydrodynamics from the Green's functions
simple we will also assume that the system is in mechanical
equilibrium i.e.\ $\bnabla p=0$, and only has gradients in the
concentration i.e.\ $\bnabla T=0$ and $\bnabla\bi u=0$.

Within the context of local equilibrium, assumed to hold at time zero
for now, we focus on the quantity $\dot{\mathsf K}_{\lambda1}$ in
equation~\eqref{complex} for component $\lambda=1$. According to
equation~\eqref{K1def},
\begin{eqnarray}
\dot{\mathsf K}_{1 1}(\bi r', t) &=& \frac{\average{\bi v_{1 1}(t)
      \delta\mathbf(\bi r'-\bi r_{1 1}(0)\mathbf)}_{\mathrm{ne}}}
  {\average{\delta(\bi r'-\bi r_{1 1}(0))}_{\mathrm{ne}}}
=
\frac{\average{N_1 \bi v_{1 1}(t)
      \delta\mathbf(\bi r'-\bi r_{1 1}(0)\mathbf)}_{\mathrm{ne}}}
  {n_1(\bi r',0)},
\label{m1}
\end{eqnarray}
where we have used that particles of the same component are
indistinguishable.  We will now evaluate $\dot{\mathsf K}_{1 1}(\bi
r', t)$ with the local equilibrium distribution $\mathcal P_{\mathrm
L}$ derived in appendix~A (following Ernst's treatment of a single
component fluid \cite{Ernst64}).  Using equation~\eqref{D2moster2} we
can write $\dot{\mathsf K}_{1 1}(\bi r',t)$ as \pagebreak[3]
\begin{eqnarray}
\fl
  \dot{\mathsf K}_{1 1}(\bi r',t)
\nonumber 
\\\fl
=
\frac{
\bigaverage{
 N_1\bi v_{1 1}(t) \delta(\bi r'-\bi r_{1 1}(0))
  \Big\{  1  + 
  \rho\displaystyle  \int\! d\bi r''  
    \left[\frac{\tilde n_1(\bi r'')}{cm_2 n}
      -  \frac{\tilde  n_2(\bi r'')}{(1-c)m_1 n}
      \right]
      (\bi r''-\bi r+\bi u\,t)\cdot\bnabla c
  \Big\} 
}_{0}
}{n_1(\bi r',0)}
\nonumber\\
\label{sixtytwoprime}
\end{eqnarray}
where $\average{A}_0$ is defined as $\int A(\Gamma)\mathcal
P_0(\Gamma) d\Gamma$ where $\mathcal P_0$ is the grand canonical
distribution but with all velocities shifted over $\bi u$, and $\tilde
n_\lambda(\bi r'')$ is the microscopic expression for the number
density of component $\lambda$ at position $\bi r''$ (see appendix~A
for details).  In appendix~B it is shown that this gives to leading
order in the gradients:
\begin{eqnarray}
\fl
\dot{\mathsf K}_{1 1}(\bi r', t) &=&
    \bi u
\label{m1result}
+\frac{\rho}{n_1}
\left[
\frac{
\average{
N_1(N_1-1)\bi v_{11}(t) \bi r_{12}(0)}
}{cm_2 n V}
-
\frac{
\average{N_1 N_2\bi v_{11}(t)\bi r_{21}(0)}
}{(1-c)m_1 n V}
\right]
\cdot
\bnabla c.
\end{eqnarray}
Using now that $\average{\bi v_{1 1}(t)\bi r_{12}(0)}=-\average{\bi
v_{11}(0)\bi r_{1 2}(t)}=\int_0^t dt'\average{\bi v_{1 1}(0)\bi v_{1
2}(t')}$ and furthermore taking the thermodynamic limit and then the
limit $t\to\infty$, we get (cf. equations~\eqref{f12} and
\eqref{fiab}) $\dot{\mathsf K}_{1 1}(\bi r', t) = \bi u +\frac{\rho^2
}{m_1 m_2 n^2}(f_{12}-f_{1\alpha\beta}) \bnabla c.$ By interchanging
$1$ and $2$, which also implies that $\bnabla c$ be replaced by
$\bnabla(1-c)=-\bnabla c$, we find the corresponding result for
component $2$, i.e.\ $\dot{\mathsf K}_{2 1}(\bi r', t) = \bi u
-\frac{\rho^2 }{m_1 m_2 n^2}(f_{12}-f_{2\alpha\beta}) \bnabla c$, so
that
\begin{equation}
\dot{\mathsf K}_{\lambda1}(\bi r',t) =
\bi u - 
(-1)^{\lambda} \frac{\rho^2}{m_1 m_2 n^2}
(f_{12}-f_{\lambda\alpha\beta})\bnabla c
\label{M1resut}
\end{equation}
Substituting this result for $\dot{\mathsf K}_{\lambda1}$ from
equation~\eqref{M1resut} into equation~\eqref{complex} and integrating
over $\bi r'$ results in
\begin{eqnarray}
\dd{n_\lambda}{t}  & = & \bnabla\cdot 
\left[
D_\lambda\bnabla n_\lambda -n_\lambda\bi u+ (-1)^\lambda \frac{\rho^2
  x_\lambda}{m_1 m_2 n}(f_{12}-f_{\lambda\alpha\beta})
\bnabla c
\right].
\label{diffusion1}
\end{eqnarray}
Using equation~\eqref{conservation}, we identify the right hand side
as $-\bnabla\cdot (n_\lambda\bi u+\bi j_\lambda/m_\lambda)$ and find
that
\begin{eqnarray}
\bi j_\lambda &=& 
-  D_\lambda\bnabla \rho_\lambda 
-(-1)^\lambda \frac{m_\lambda\rho^2x_\lambda}{m_1 m_2 n}(f_{12}-f_{\lambda\alpha\beta})
\bnabla c
\nonumber
\\
&=&
 (-1)^\lambda  \frac{m_\lambda\rho^2}{m_1 m_2 n}\left[D_\lambda- x_\lambda
(f_{12}-f_{\lambda\alpha\beta})
\right]\bnabla c.
\label{j1res}
\end{eqnarray}
where we have used equation~\eqref{rhociso} to write $\bnabla\rho_1$
at constant pressure and temperature (which is the case here) as
$[\rho+c(\partial\rho/\partial c)_{Tp}]\bnabla c= \rho^2/(m_2
n)\bnabla c$, and similarly $\bnabla \rho_2=-\rho^2/(m_1 n)\bnabla c$.
Since $\bi j_\lambda=(-1)^\lambda\rho D \bnabla c$ (by definition), we
see from equation~\eqref{j1res} that the mutual diffusion constant is
given by
\begin{eqnarray}
  D &=& \frac{m_\lambda\rho}{m_1 m_2
n}[D_\lambda-x_\lambda(f_{12}-f_{\lambda\alpha\beta})]
=
- \frac{\rho^2}{m_1m_2n^2} f_{12}
.
\label{finally}
\end{eqnarray}
where in the last step we used equation~\eqref{relation1} for the case
$\lambda=1$ and equation~\eqref{relation2} for the case $\lambda=2$.
The Green's function expression for the mutual diffusion in
equation~\eqref{finally} is therefore identical to the Green-Kubo
result in equation~\eqref{30prime}.

So far in this section the system was assumed to be in local
equilibrium at time $t=0$. We are really interested in systems that
are initially far from equilibrium. If left unperturbed, such a system
would still reach local equilibrium, after a time that we shall denote
by $t_L$ and which depends on the details of the initial condition
(with no general way to determine it other than through measurement).

A natural question is therefore whether the Gaussian part of the
Green's function for such a far from equilibrium system is in
agreement with hydrodynamics for times $t\gg t_L$ as well. We will now
argue that this is indeed the case.

For this, we start from equation~\eqref{complex} for $\partial
n_\lambda/\partial t$. That equation was obtained by using 1) the
Gaussian approximation for the Green's function, and 2) that $t$ is
large enough for $\dot{\mathsf K}_{\lambda 2}$ to have reached its
long time value $2D_\lambda$. Since the assumption of local
equilibrium was not needed for the derivation of
equation~\eqref{complex}, it holds also in cases where the system
starts far from equilibrium.  To obtain hydrodynamics from
equation~\eqref{complex}, we calculated $\dot{\mathsf K}_{\lambda 1}$
assuming that the system was in local equilibrium at time $0$. Now we
would like to do the same in situations where local equilibrium is
established only later, at time $t_L$. The local equilibrium condition
at time $t_L$ can be used by writing $\dot{\mathsf K}_{\lambda 1}$ as
(cf.~equation~\eqref{K1def})
\begin{equation}
\fl
  \dot{\mathsf K}_{\lambda 1} = 
\average{\bi v_{\lambda1}(t)}_{\lambda\bi r'}
= \int d\bi r'' 
\frac{
\average{
\bi v_{\lambda1}(t)\delta(\bi r''-\bi r_{\lambda1}(t_L))
}
_{\lambda\bi r'}
}
{
\average{\delta(\bi r''-\bi r_{\lambda1}(t_L))}
}
\average{\delta(\bi r''-\bi r_{\lambda1}(t_L))}
\label{link}
\end{equation}
Assuming now that all microscopic correlations between quantities at
time $0$ and at later times have died out before $t_L$, the condition
that $\bi r_{\lambda1}(0)=\bi r'$ in $\average{ \bi
v_{\lambda1}(t)\delta(\bi r''-\bi r_{\lambda1}(t_L)) }_{\lambda\bi
r'}$ should not matter, i.e.
\begin{equation}
\average{\bi v_{\lambda1}(t)\delta(\bi r''-\bi r_{\lambda1}(t_L))}
_{\lambda\bi r'}
\approx
\average{
\bi v_{\lambda1}(t)\delta(\bi r''-\bi r_{\lambda1}(t_L))
}
\end{equation}
Using this in equation~\eqref{link} gives
\begin{equation}
  \dot{\mathsf K}_{\lambda 1} = 
 \int d\bi r'' 
\average{\bi v_{\lambda1}(t)}
_{\bi r_{\lambda1}(t_L)=\bi r''}
\average{\delta(\bi r''-\bi r_{\lambda1}(t_L))}
\label{approxlink}
\end{equation}
where the subscript appended to the average $\average{\bi
v_{\lambda1}(t) }_{\bi r_{\lambda1}(t_L)=\bi r''} $ indicates that it
is taken only over all configuration for which $\bi
r_{\lambda1}(t_L)=\bi r''$.  The quantity $\average{\bi
v_{\lambda1}(t) }_{\bi r_{\lambda1}(t_L)=\bi r''} $ can be calculated
using local equilibrium, since the condition involves time $t_L$, at
which time we assume local equilibrium to have been established. In
fact, except for the time shift over $t_L$, it is precisely the same
quantity $\dot{\mathsf K}_{\lambda1}$ that was calculated for the case
where local equilibrium was present at time zero, with the result
\eqref{M1resut}, so that
\begin{equation}
\average{\bi v_{\lambda1}(t)}
_{\bi r_{\lambda1}(t_L)=\bi r''}
=
\bi u - 
(-1)^{\lambda} \frac{\rho^2}{m_1 m_2 n^2}
(f_{12}-f_{\lambda\alpha\beta})\bnabla c.
\end{equation}
This expression for $\average{\bi v_{\lambda1}(t)} _{\bi
r_{\lambda1}(t_L)=\bi r''} $ is independent of $\bi r''$, so that
after its substitution into equation~\eqref{approxlink}, the
integration over $\bi r''$ can be performed giving
\begin{equation}
  \dot{\mathsf K}_{\lambda 1} = 
\bi u - 
(-1)^{\lambda} \frac{\rho^2}{m_1 m_2 n^2}
(f_{12}-f_{\lambda\alpha\beta})\bnabla c.
\label{final}
\end{equation}
This is identical to the case where local equilibrium was established
at time zero, i.e.\ equation~\eqref{M1resut}. Substituting
equation~\eqref{final} in equation~\eqref{complex} would thus lead, as
before, to the hydrodynamic equation \eqref{diffusion1}, with the
right value for the mutual diffusion constant as given by
equation~\eqref{finally}.

In conclusion, the Gaussian approximation to the Green's function for
a far from equilibrium initial situation gives hydrodynamic
behaviour for long times $t>t_L$ as long as local equilibrium is
established at some finite time $t_L$ (not necessarily zero).

\section{Discussion}
\label{conclusions}

In this paper we have shown that for binary diffusion in an isotopic
or ideal mixture, the Green's function theory is able to handle long
time behaviour.

We remark that the very short time behaviour was discussed in a
previous paper \cite{VanZonCohen05c}, using a recent theorem regarding
particle displacements and their cumulants under the condition that
the initial velocity distribution is Gaussian while the initial
position distribution may be arbitrary.

The long time behaviour was checked here by expanding in gradients
and using a local equilibrium assumption for long times. The latter
assumption is natural because it is generally expected that local
equilibrium is a condition for hydrodynamic behaviour.

Given the form of the Green's functions formula in
equation~\eqref{2.1}, in which the density of component $\lambda$ at
time $t$ is expressed solely in terms of that same density at time
zero, it may be surprising that the Green's functions do not produce
only self-diffusion. In fact the Green's functions reproduce mutual
diffusion with the correct (Green-Kubo) expression for the mutual
diffusion constant (see equation~\eqref{realDM}) with both the self
(Darken-Hartley-Crank) terms and the necessary cross terms
$f_{\lambda\alpha\beta}$ and $f_{12}$. Note that the cross terms found
in equations~\eqref{finally} arose from the local equilibrium
assumption applied to the drift terms $\dot{\mathsf K}_{\lambda1}(\bi
r',t)$, whereas the self-terms come from the rather simple
approximation to the variance $\dot{\mathsf K}_{\lambda2}(\bi
r',t)\approx 2D_\lambda$. We can therefore conclude that although the
Green's functions seem to treat the system as self-diffusion, the
sensitivity of the drift terms to local equilibrium restores the true
mutual and hydrodynamic character of the binary diffusion.

We note that the Green's function formalism presented here can be
straightforwardly generalized to general classical multi-component
mixtures with a non-equilibrium initial distribution, since the
equations of the Green's functions,
equations~\eqref{2.1}--\eqref{somecorrections}, do not rely on the
fact that there are only two components, nor on the fact that they are
isotopic or ideal. It was mainly for the discussion of the approach to
hydrodynamics in section~\ref{long-time-behavior} that we restricted
ourselves to isotopic and ideal binary mixtures in order not to make
the manipulations overly complicated.

We will end with a few questions to be addressed in the future.

Although we expect that the Green's functions will also be able to
reproduce hydrodynamics at long times in other systems, especially in
general mixtures, this is at present an open problem.

The Green's function formalism presented in
section~\ref{greens-funct-meth} only describes mass
transport. Analogous to the treatment in
references~\cite{KincaidCohen02a,KincaidCohen02b}, it would be
desirable to extend the formalism presented here to include local
currents and energy densities.

The \emph{intermediate} time behaviour of the Green's functions for
mixtures --- expected to be more relevant for nanoscale systems --- is
to be checked in computer simulations. Here, the correction terms
$\mathsf B_{\lambda n}$ in equation~\eqref{Gspat} will become
important. If the situation is similar to that of the heat pulse
studied in
references~\cite{KincaidCohen02a,KincaidCohen02b,CohenKincaid02},
taking the Gaussian approximation alone would turn out not to suffice
for the correct description of the behaviour of the system, but that
including just a few correction terms, such as in
equation~\eqref{somecorrections}, should be enough to get good
agreement with the simulations on the picosecond scale. Note that
these correction terms represent non-Gaussian behaviour of the
particle displacement; thus it is expected that non-Gaussian behaviour
occurs on the time scale in between the initial infinitesimal and the
long time, hydrodynamic behaviour.

These issues are currently under investigation.

\ack

We would like to thank Prof.\ J.\ M.\ Kincaid for useful discussions.
We also acknowledge the support of the Office of Basic Energy Sciences
of the US Department of Energy (grant no.\ DE-FG-02-88-ER13847) and of
the National Science Foundation (grant no. PHY-0501315).

\appendix
\renewcommand{\thesection}{\Alph{section}}

\section{Local equilibrium}
\label{app1}

In this appendix we will formulate the local equilibrium distribution
in an expansion in gradients following the method of Ernst
\cite{Ernst64} for a one-component fluid. For a one component system
Ernst derived an expression for the (grand canonical-like) local
equilibrium phase space distribution function at time $t$ expanded in
gradients of the local thermodynamic fields $T$ (temperature), $\bi u$
(fluid velocity) and $\mu$ (chemical potential), at a position $\bi r$
and time $t$ at which one wishes to calculate a local quantity. We
will use a similar derivation here for a two component mixture, in
which case there are two chemical potentials.  As in the main text, we
will consider a case with only gradients in the chemical potentials
$\mu_\lambda$, while $\bnabla p=0$, $\bnabla T=0$ and $\bnabla \bi
u=0$.

The local equilibrium distribution will be defined in analogy with the
usual grand canonical equilibrium distribution function at temperature
$T$, chemical potentials $\mu_1$ and $\mu_2$ and an average fluid
velocity $\bi u=0$, given by
\begin{equation}
\mathcal  P(\Gamma) 
  = 
  \frac{\exp\{   - [H(\Gamma) -\sum_{\lambda =1}^2\mu_\lambda N_\lambda]/k_BT\}
}{N_1!N_2!h^{3N}Z}
,
\label{Deqdef}
\end{equation} 
where the Hamiltonian is defined in equation~\eqref{Hamiltonian} and
the normalization $Z = \sum_{N_1}\sum_{N_2} \int\!d\Gamma$ $\exp\{ -[
H(\Gamma) -\mu_1 N_1 -\mu_2 N_2]/k_BT\}$ $/(N_1!N_2!h^{3N}) $.  For
local equilibrium, we suppose that at a certain instant, taken to be
time $0$, one can define a uniform fluid velocity $\bi u$ and spatial
varying chemical potentials of the two components, $\mu_1(\bi r)$ and
$\mu_2(\bi r)$, while $p$ and $T$ are constant.  The local equilibrium
distribution is then
\begin{eqnarray}
 \mathcal P_{\mathrm L}\mathbf(\Gamma|\mu_{\lambda}) 
   \equiv 
  \frac{
  \exp\{-[ H_p(\Gamma)-\int d\bi r'\,
    \sum_{\lambda =1}^2 \mu_\lambda(\bi r') \tilde n_\lambda(\bi r') 
] /k_BT\}
}{N_1!N_2!h^{3N}Z_{L}},
\label{Dloc}
\end{eqnarray}
where the normalization $Z_\lambda$ is given by
\begin{eqnarray}
 Z_{\mathrm L}
   = 
  \sum_{N_1=0}^\infty\sum_{N_2=0}^\infty
\int\!d\Gamma
    \frac{
 \exp\{-[ H_p(\Gamma)-\int d\bi r'\sum_{\lambda =1}^2 
 \mu_\lambda(\bi r') \tilde n_\lambda(\bi r') 
]/k_BT
\}
}{N_1!N_2!h^{3N}}
.
\label{Zloc}
\end{eqnarray}
Furthermore, in equations~\eqref{Dloc} and \eqref{Zloc} the phase
functions corresponding to the local number densities and the total
peculiar energy are (cf.\ equations~\eqref{1.1} and
\eqref{Hamiltonian})
\begin{eqnarray}
  \tilde n_\lambda(\bi r') 
  & \equiv & 
  \:\:\:\sum_{ i=1}^{N_\lambda}\:\: \delta(\bi r'-\bi r_{\lambda  i})
\label{n1def}
\\
  H_p(\Gamma) 
  & \equiv & 
 \sum_{\lambda =1}^{2} \sum_{ i=1}^{N_\lambda} 
    \Big[
      \case12m_\lambda|\bi v_{\lambda  i} - \bi u|^2 
    + \sum_{\nu=1}^{2}{\sum_{j=1}^{N_\nu}}{\Big.}' 
  \case{1}{2}\varphi_{\lambda\nu}
(|\bi r_{\lambda  i}-\bi r_{\nu j}|)
      \Big].
\label{edef}
\end{eqnarray}
We note that the requirement of constant pressure puts, through the
Gibbs-Duhem relation \eqref{GD}, a restriction on the functions
$\mu_\lambda(\bi r)$:
\begin{equation}
 \bnabla \mu_2 = -{n_1}/{n_2}\bnabla
\mu_1 
\label{number}
.
\end{equation}

Following Ernst we expand the distribution function as
\begin{equation}
    \mathcal P_L(\Gamma|\mu_\lambda) 
    = 
    \mathcal P_0(\Gamma|\mu_{\lambda}) 
      + \mathcal P_1(\Gamma|\mu_{\lambda}) +\ldots
\label{expand}
\end{equation}
Here $\mu_{\lambda}=\mu_{\lambda}(\bi r,t)$ with $\bi r$ a specific
position around which local quantities are to be
evaluated. Furthermore, the leading term $\mathcal
P_0(\Gamma|\mu_{\lambda}) =
P_L(\Gamma|\mu_\lambda)\big|_{\mu_\lambda=\mu_\lambda(r,t)} $, which
is a distribution of the form of a grand canonical equilibrium
distribution in equation~\eqref{Deqdef} at the (fixed) values
$\mu_\lambda=\mu_{\lambda}(\bi r, t)$ and in terms of peculiar
velocities i.e.\ with $H$ replaced by $H_p$.

Expanding equation~\eqref{Dloc} around
$\mu_\lambda\equiv\mu_\lambda(r,t)$, we find for the next-to-leading
term in equation~\eqref{expand}
\begin{eqnarray}
\fl
  \mathcal P_1(\Gamma|\mu_{\lambda}) 
  = 
    \int d\bi r'  \sum_{\lambda =1}^2
    \left.\frac{\delta
     \mathcal P_{\mathrm L}(\Gamma|\mu_{\lambda})}{\delta\mu_\lambda(\bi r')}\right|_{\mu_{\lambda}\equiv\mu_{\lambda}(\bi r, t)}
  \left[
  (\bi r'-\bi r)\cdot
  \bnabla \mu_\lambda(\bi r, t) - t\dd{}{t}\mu_\lambda(\bi r, t)
\right]
\label{formal}
\end{eqnarray}
where we used that $\mu_\lambda(\bi r',0) = \mu_\lambda(\bi r, t)
+(\bi r'-\bi r)\cdot\bnabla \mu_\lambda(\bi r, t) -
t(\partial/\partial t)\mu_\lambda(\bi r, t) +\ldots$ .  The functional
derivatives are, using equation~\eqref{Dloc},
\begin{eqnarray}
  \frac{\delta \mathcal P_{\mathrm L}(\Gamma|\mu_{\lambda})}{\delta\mu_\lambda(\bi r')} 
  & = &  
  \mathcal P_{\mathrm L}(\Gamma|\mu_{\lambda})
   \Bigg[
     \frac{\tilde n_\lambda(\bi r') }{k_BT} - \frac{\delta
      \ln Z_{\mathrm L}}{\delta\mu_\lambda(\bi r')}
   \Bigg]
\label{functional}
,
\end{eqnarray}
The term $\delta \ln Z_{\mathrm L}/\delta\mu_\lambda(\bi r')$ in
equation~\eqref{functional} can be shown to be the average of $\tilde
n_\lambda(\bi r')/k_BT$ taken with $\mathcal P_{\mathrm L}$, so, as it
occurs in equation~\eqref{formal},
\begin{eqnarray}
  \left.
  \frac{\delta \mathcal P_{\mathrm L}(\Gamma|\mu_{\lambda})}{\delta\mu_\lambda(\bi r')}
   \right|_{\mu_{\lambda}\equiv\mu_{\lambda}(\bi r, t)}
  & = &  
  \frac{\mathcal P_{\mathrm L}(\Gamma|\mu_{\lambda})}{k_BT}
   \big[\tilde n_\lambda(\bi r') -\average{\tilde n_\lambda(\bi r')}_{L}\big]
   \bigg|_{\mu_{\lambda}\equiv\mu_{\lambda}(\bi r, t)}
\nonumber\\
  & = &  
  \frac{\mathcal P_0(\Gamma|\mu_\lambda)}{k_BT}
   [\tilde n_\lambda(\bi r') 
   -n_\lambda]
\label{derivatives}
,
\end{eqnarray}
where we used that the average $\average{}_{\mathrm
L}\approx\average{}_0$ is translation invariant so $\average{\tilde
n_\lambda(\bi r')}_0=n_\lambda$.  We substitute
equation~\eqref{derivatives} into equation~\eqref{formal} and obtain
\begin{eqnarray}
\fl
  \mathcal P_1(\Gamma|\mu_{\lambda}) \:=
  & 
     \frac{\mathcal P_0(\Gamma|\mu_{\lambda})}{k_BT} \!\int\! d\bi r'
  &
  \sum_{\lambda=1}^2
[\tilde n_\lambda(\bi r') 
  -n_\lambda
  ]
  [
  (\bi r'-\bi r)\cdot\bnabla\mu_\lambda
- t\dd{}{t}\mu_\lambda
  ]
\label{D1first}
.
\end{eqnarray}
In equation~\eqref{D1first}, we are only interested in
$\partial\mu_\lambda/\partial t$ to leading order in the gradients.
This is given by the Euler equations which are just
equations~\eqref{1}--\eqref{4} without dissipation terms i.e.\
$D\rho/Dt=Dc/Dt=0$. The quantities $\mu_\lambda$ are just functions of
$\rho$ and $c$ (and of $T$, which is constant), so that also
$D\mu_\lambda/Dt=0$ and equation~\eqref{D1first} becomes
\begin{eqnarray}
  \mathcal P_1(\Gamma|\mu_{\lambda}) &= 
  &
   \frac{\mathcal P_0(\Gamma|\mu_{\lambda})}{k_BT}
\int\! d\bi r' 
\sum_{\lambda=1}^2
  [\tilde n_\lambda(\bi r') 
    -n_\lambda
   ]\bnabla\mu_\lambda
    \cdot (\bi r'-\bi r+\bi u  t).
\label{D2first}
\end{eqnarray}

As in the main text we use instead of $\mu_1$ the quantity $ \mu =
{\mu_1}/{m_1}-{\mu_2}/{m_2} $.  with the help of the Gibbs-Duhem
relation \eqref{number}, which gives $\bnabla \mu_2 =
-({n_1}/{n_2})\bnabla \mu_1 $, we find $\bnabla\mu_1 = m_1
(1-c)\bnabla\mu$ and $\bnabla\mu_2 = -m_2 c\bnabla\mu$.  Substituting
this into equation~\eqref{D2first} and using equation~\eqref{bmuciso}
gives the result used in the main text:
\begin{eqnarray}
\fl
  \mathcal P_1(\Gamma|\mu_{\lambda}) &=&
\frac{\mathcal P_0(\Gamma|\mu_{\lambda})}{k_BT}
  \int d\bi r' [(1-c)m_1\tilde n_1(\bi r') -
     c m_2 \tilde n_2(\bi r') 
    ]
    \bnabla \mu
  \cdot(\bi r'-\bi r+\bi u  t)
\nonumber
\\\fl
  &
=
& \ddc{\mu}{c}{Tp}
     \frac{\mathcal P_0(\Gamma|\mu_{\lambda})}{k_BT}
   \int\! d\bi r'\,
  [(1-c) m_1\tilde n_1(\bi r') 
    - c m_2\tilde n_2(\bi r') 
    ]
(\bi r'-\bi r+\bi u t)
\cdot
  \bnabla c
\nonumber\\\fl
  &= &
    \mathcal P_0(\Gamma|\mu_{\lambda})\,
\rho \mathcal   \int d\bi r'
  \left[\frac{\tilde n_1(\bi r')}{cm_2 n  }
    - \frac{\tilde n_2(\bi r')}{(1-c)m_1 n }
    \right]
 [\bi r'-\bi r+\bi u t]
\cdot
  \bnabla c.
\label{D2moster2}
\end{eqnarray}

\section{Derivation of equation~\eqref{m1result}}
\label{app2}

The average $\average{}_{0}$ in equation~\eqref{sixtytwoprime} is over
a variant of the grand canonical ensemble where each particle has an
additional velocity $\bi u$ compared to the normal grand canonical
equilibrium ensemble. Since each particle has the same velocity shift,
this does not change the relative motion of the particles. Hence, we
can replace $\average{}_{0}$ in the above expression by a real
canonical ($\bi u=0$) ensemble average $\average{}$, if we replace
$\bi v_{1 1}(t)$ by $\bi u+\bi v_{1 1}(t)$ in
equation~\eqref{sixtytwoprime}:
\begin{eqnarray}
\fl
  \dot{\mathsf M}_{1 1}(\bi r',t) & = &
\frac1{n_1(\bi r',0)}\Big\langle
N_1 \delta(\bi r'-\bi r_{11}(0))
[\bi u+\bi v_{11}(t)]
\nonumber
\\\fl&&\qquad\:\times
  \Big\{  1  + 
    \rho \int d\bi r''  
    \left[\frac{\tilde n_1(\bi r'') }{cm_2 n}
      - \frac{\tilde n_2(\bi r'')}{(1-c)m_1 n}
      \right]
      (\bi r''-\bi r+\bi u\,t)\cdot \bnabla c
  \Big\} 
\Big\rangle.
\end{eqnarray}
The part of the average proportional to $\bi u$ is
precisely equal to $n_1(\bi r',0)$, so
\begin{eqnarray}
\fl \dot{\mathsf M}_{1 1}(\bi r',t) & = & \bi u +
\frac{\langle
N_1 \delta(\bi r'-\bi r_{1 1}(0))
\bi v_{1 1}(t)
\rangle
}{n_1(\bi r',0)}
\\\fl
&+&
\rho \bigg\langle
N_1 \delta(\bi r'-\bi r_{1 1}(0))
\bi v_{1 1}(t)
     \int d\bi r''  
    \Big[\frac{\tilde n_1(\bi r'') }{cm_2 n}
      - \frac{\tilde n_2(\bi r'')} {(1-c)m_1 n}
      \Big]
\bigg\rangle
\frac{(\bi r''-\bi r+\bi u\,t)\cdot
\bnabla c}{n_1(\bi r',0)}
\nonumber
\end{eqnarray}
To simplify this expression further we will apply the following trick,
based on the fact that $\average{\bi v_{11}(t)}=0$, several times.
The term $\langle N_1 \delta(\bi r'-\bi r_{1 1}(0)) \bi v_{1 1}(t)
\rangle$ can be shown to vanish as follows. Its only spatial
dependence is through $\bi r'$ but because of translation invariance
it cannot really depend on $\bi r'$ so we can integrate over $\bi r'$
and divide by $V$. This leaves us with $\average{N_1 \bi v_{1
1}(t)}/V$.  Then $\bi v_{1 1}(t)$ can be replaced by $\bi v_{1 1}(0)$
because both the number of particles and the grand canonical
distribution function are invariant under time evolution and the
average of it is then seen to be zero, i.e.\ $\average{N_1 \bi v_{1
1}(0)}/V=0$. Thus
\begin{eqnarray}
\fl
\dot{\mathsf M}_{1 1} (\bi r',t)  &=&  \bi u +
\label{T1}
    \frac{\rho}{n_1} \int\! d\bi r'' \, \bigaverage{
N_1 \delta(\bi r'-\bi r_{1 1}(0))
\bi v_{1 1}(t)
    \left[\frac{\tilde n_1(\bi r'')}{cm_2 n}
      - \frac{\tilde n_2(\bi r'')}{(1-c)m_1 n} 
      \right]
}
\nonumber\\\fl&&\qquad\times
     (\bi r''-\bi r+\bi u\,t)
\cdot
\bnabla c,
\label{ares}
\end{eqnarray}
where we also replaced $n_1(\bi r',0)$ by $n_1=n_1(\bi r, t)$, since
the difference can be expressed in terms of gradients and would
contribute only to second order in the gradients in
equation~\eqref{ares}.  The term proportional to $(-\bi r+\bi u t)$
can be integrated over $\bi r''$. Using $\int d\bi r'' \tilde
n_\lambda(\bi r'') = N_\lambda$, this yields factors of $N_1^2$ and
$N_1 N_2$ times $\bi v_{1 1}(t)\delta(\bi r'-\bi r_{1 1}(0))$. By the
same reasoning as above this is zero, and we are left with
\begin{eqnarray}
\fl
 \dot{\mathsf M}_{1 1}(\bi r',t) &=& \bi u +
\label{T2}
   \frac{\rho}{n_1} \int d\bi r''  
\bigaverage{\!
N_1\delta\mathbf(\bi r'-\bi r_{1 1}(0)\mathbf)
\bi v_{1 1}(t)
    \left[\frac{\tilde n_1(\bi r'')}{cm_2 n}
      - \frac{\tilde n_2(\bi r'') }{(1-c)m_1 n}
      \right]
\!
}
\bi r''\!\cdot\! \bnabla c.
\end{eqnarray}
The remaining integral over $\bi r''$ can be performed by using the
explicit forms of $\tilde n_1(\bi r'')$ and $\tilde n_2(\bi r'')$ in
equation~\eqref{n1def}. This gives:
\pagebreak[3]
\begin{eqnarray}
\fl
\dot{\mathsf M}_{1 1}(\bi r',t) 
\nonumber
\\\fl{}  = \bi u
+\frac{\rho}{n_1} \average{
N_1
\bi v_{11}(t)
\delta(\bi r'-\bi r_{1 1}(0))
    \left[\frac{\bi r'}{cm_2 n}  +
\sum_{ i=2}^{N_1}\frac{\bi r_{1 i}(0)}{cm_2 n}  
      - 
\sum_{ i'=1}^{N_2}
\frac{
 \bi r_{2 i'}(0)}{(1-c)m_1 n}
      \right]
}
\cdot 
\bnabla c.
\nonumber
\end{eqnarray}
Here we have written out the term $ i=1$ explicitly. This term is in
fact zero: once the factor of $\bi r'$ has been taken out of the
average, we can use the above reasoning that the remainder by
translation invariance does not depend on $\bi r'$ and can be
integrated~over $\bi r'$, after which we have an integral of a time
invariant expression times $\bi v_{1 1}(t)$, which is zero.  Using
also that the particles of each component are indistinguishable, we
find
\begin{eqnarray}
\dot{\mathsf M}_{1 1}(\bi r',t) & = &  \bi u
+
\frac{\rho}{n_1}\Bigg[
\frac{
\average{
\delta(\bi r'-\bi r_{11}(0))N_1(N_1-1)\bi
v_{11}(t) \bi r_{12}(0)}
}{cm_2 n}
\nonumber\\&&\qquad
-
\frac{ 
\average{\delta(\bi r'-\bi r_{11}(0))N_1 N_2\bi
  v_{11}(t)\bi r_{21}(0)}
}{(1-c)m_1 n}
\Bigg]
\cdot
\bnabla c.
\end{eqnarray}
This whole expressions is independent of $\bi r'$. It may seem to
change under translation over some fixed vector $\bi s$ as this would
change $\bi r_{\lambda i}$ to $\bi r_{\lambda i}+\bi s$. However, the
prefactor in front of $\bi s$ that results is itself independent of
$\bi r'$, can therefore be integrated over $\bi r'$, and is then zero
because of the factor $\bi v_{1 1}(t)$. Hence the expression is
translation invariant and may be integrated over $\bi r'$ and divided
by $V$.  This then gives equation~\eqref{m1result}.

\end{document}